\documentclass[prd,nofootinbib,amsfonts,notitlepage]{revtex4-2}
\usepackage[utf8]{inputenc}
\usepackage[T1]{fontenc}
\usepackage{hyperref}
\usepackage{graphicx,bm,amsmath,color}
\usepackage{bbold}
\usepackage{wasysym}
\usepackage{verbatim}
\usepackage{amssymb}
\usepackage{epstopdf}
\usepackage{animate}
\usepackage{xmpmulti}
\usepackage{centernot}
\usepackage{multirow}
\usepackage{tikz}
\usepackage{graphicx}
\usepackage{float}
\usepackage{mathtools}
\usepackage{comment}
\usetikzlibrary{arrows}
\usetikzlibrary{decorations.pathreplacing}

\makeatletter

\newcommand{\be}{\begin{equation}}
\newcommand{\ee}{\end{equation}}
\newcommand{\beq}{\begin{eqnarray}}
\newcommand{\eeq}{\end{eqnarray}}
\newcommand{\ba}{\begin{align}}
\newcommand{\ea}{\end{align}}

\begin{document}

\title{Exploring black hole mechanics in cotangent bundle geometries}
\author{José Javier Relancio}
\affiliation{Dipartimento di Fisica ``Ettore Pancini'', Università di Napoli Federico II, Napoli 80125, Italy;\\
INFN, Sezione di Napoli, Napli 80125, Italy;\\
Departamento de F\'{\i}sica Te\'orica and Centro de Astropartículas y Física de Altas Energías (CAPA),
Universidad de Zaragoza, Zaragoza 50009, Spain}
\email{relancio@unizar.es}

\author{Stefano Liberati}
\email{liberati@sissa.it}
\affiliation{SISSA, Via Bonomea 265, 34136 Trieste, Italy and INFN, Sezione di Trieste;\\ IFPU - Institute for Fundamental Physics of the Universe, Via Beirut 2, 34014 Trieste, Italy}

\begin{abstract}
The classical and continuum limit of a quantum gravitational setting could lead, at mesoscopic regimes, to a very different notion of geometry w.r.t. the pseudo-Riemannian one of special and general relativity. A possible way to characterize this modified space-time notion is by a momentum dependent metric, in such a way that particles with different energies could probe different spacetimes. Indeed, doubly special relativity theories, deforming the special relativistic kinematics while maintaining a relativity principle, have been understood within a geometrical context, by considering a curved momentum space. The extension of these momentum spaces to curved spacetimes and its possible phenomenological implications have been recently investigated. Following this line of research, we address here the first two laws of black holes thermodynamics in the context of a cotangent bundle metric, depending on both momentum and space-time coordinates, compatible with the relativistic deformed kinematics of doubly special relativity. 

\end{abstract}

\maketitle

\section{Introduction}
The quest for a quantum gravity theory (QGT) able to conjugate the two pillars of modern physics, quantum field theory (QFT) and general relativity (GR), has been the object of intense scrutiny for several decades by now. Indeed, while QFT and GR have been proved to describe respectively the interaction of elementary particles and astrophysical objects with extraordinary accuracy, we still need a QGT in order to have predictability in extreme situations, like the very early universe or the interior of black holes. 
However such a theory, e.g.~something able to describe the interaction and propagation high-energetic small particles on a quantum spacetime, is still missing.
%One of the incompatibilities arises in the fundamental notion of spacetime, which is an static frame in QFT and a dynamical variable in GR. 

Nonetheless, in the past decades important advancements towards this aim have been accomplished leading to the construction of remarkable theoretical frameworks such as string theory~\cite{Mukhi:2011zz,Aharony:1999ks,Dienes:1996du}, loop quantum gravity~\cite{Sahlmann:2010zf,Dupuis:2012yw}, causal dynamical triangulations \cite{Loll_2019} or causal set theory~\cite{Wallden:2013kka,Wallden:2010sh,Henson:2006kf}.  In most of these theories, a minimum length appears~\cite{Gross:1987ar,Amati:1988tn,Garay1995},  normally associated with the Planck length $\ell_P \sim 1.6\times 10^{-33}$\,cm (corresponding to the energy scale $M_P \sim 1.22\times 10^{19}$\,GeV). This length is supposed to separate somehow the classical and ``quantum'' regimes of spacetime, a transition that might lead to novel scenarios and possible observable implications. 

Unfortunately, the aforementioned theories are not yet able to provide definitive predictions concerning the phenomenology associated to this transition (albeit several proposal in this sense are present in the extant literature). Given this state of affairs, in recent times it has been considered an alternative approach: instead of considering a top-down investigation based on some candidate QGT, one can take a complementary bottom-up procedure based on postulating different kind of modifications of the space-time geometry at mesoscopic regimes, i.e., between the observed scales and the Planck one. Such departures from the standard structure of a classical/continuum spacetime can be on one side related to different quantum gravity scenarios, on the other side they are often amenable to phenomenological testing/constraining, so they can potentially provide important clues about the features that a QGT should have in order to be compatible with present observations at energies below the Planck scale. 

A very commonly considered scenario is the one in which some local space-time symmetries, such as Poincar\'e invariance, are modified at very high energies. 
In particular, a large body of research has been dedicated to the possible phenomenology associated to departures from local Lorentz invariance, focusing mostly on modifications of the dispersion relations for elementary particles of the form $E^2=p^2+m^2+f(E,p,M_P)$. Most interestingly, using current experiments and observations, several significant constraints can be put on the strength of the above introduced Lorentz violating term $f(E,p,M_P)$ (see e.g.~\cite{Mattingly:2005re,Liberati:2013xla}).  

Let us stress here that such dispersion relations can be associated alternatively to broken or deformed local Lorentz invariance.
The first scenario, in which Lorentz symmetry (more precisely the relativity principle at its base) is broken, can be considered by studying for example extensions of the standard model of particle physics including terms allowing for Lorentz invariance violation (LIV)~\cite{Colladay:1998fq,Kostelecky:2008ts,Mattingly:2005re,Liberati:2013xla}. Similar studies can be carried on also in the gravitational sector by introducing a suitable dynamical aether vector field, see e.g.~\cite{Jacobson:2000xp,Horava:2009uw,Liberati:2013xla}. Alternatively, one can contemplate different scenarios in which the relativity principle is preserved but its implementation is deformed in the sense that the action of the boosts generator on momenta is modified, so to leave invariant some form of the aforementioned modified dispersion relation. This approach takes the name of doubly special relativity (DSR)~\cite{AmelinoCamelia:2008qg}. 

%The main modification of LIV scenarios with respect to the kinematics of special relativity (SR) is a modified dispersion relation. Usually, the quadratic expression of SR includes in this setting new terms proportional to new powers of momentum and inversely proportional to a high-energy scale. In this kind of theories, there is no longer the relativity principle that characterize SR, having a privileged observer (usually considered as the one who observes the cosmic microwave background isotropic). Geometrically such theories require the introduction of an aether vector field providing the preferred frame associated to LIV \cite{Jacobson:2000xp,Horava:2009uw}.

An important characteristic of DSR theories is that, besides a possible deformation of the dispersion relation, a deformed (non-additive) composition law for momenta is necessary. 
While the first ingredient (the dispersion relation) can be even unmodified, i.e., have the same expression of SR~\footnote{This happens in the so-called ``classical basis'' of $\kappa$-Poincaré~\cite{Borowiec2010}}, the last one must be always deformed, being then the crucial difference with respect to SR kinematics. 
%Moreover, there is some Lorentz transformations, in the one- and two-particle sector, preserving a relativity principle.  

It was ``a posteriori'' recognized that also in this framework a geometrical interpretation of the deformed symmetries is possible through the introduction of a velocity or momentum dependent spacetime. The most common examples of this kind of geometries are Finsler and Hamilton geometries respectively~\cite{miron2001geometry}, and indeed within these frameworks, the aforementioned departures from standard SR have been considered~\cite{Girelli:2006fw,Amelino-Camelia:2014rga,Letizia:2016lew}~\footnote{Albeit, it is perhaps worth stressing that also LIV scenarios can sometime be described via the same kind of momentum/velocity dependent geometries, see e.g.~\cite{Kostelecky:2011qz,Barcelo:2001cp,Weinfurtner:2006wt,Hasse:2019zqi,Stavrinos:2016xyg}.}.
Similarly, momentum dependent geometries in this DSR framework have also been considered in~\cite{Barcaroli:2016yrl,Barcaroli:2015xda,Barcaroli:2017gvg}. 

An open problem common to the above mentioned approaches is related to the apparent impossibility to properly take into account the role of the momenta composition law. Nonetheless, in a parallel stream of research it was shown in~\cite{Carmona:2019fwf} that all the ingredients of a relativistic deformed kinematics (modified dispersion relation and composition law) can be obtained from a maximally symmetric momentum space. In particular,  $\kappa$-Poincaré kinematics can be obtained identifying the isometries (translations and Lorentz isometries) and the squared distance of the metric with the deformed composition law, deformed Lorentz transformations and deformed dispersion relation respectively (the last two facts were previously contemplated in Refs.~\cite{AmelinoCamelia:2011bm,Lobo:2016blj}). 

In~\cite{Relancio:2020zok,Relancio:2020rys,Relancio:2020mpa,Pfeifer:2021tas,Relancio:2021asx} the proposal of~\cite{Carmona:2019fwf} was generalized so allowing the metric to describe a curved spacetime, so leading to a geometry in the cotangent bundle depending on all the phase-space variables. This scheme is a generalization of the aforementioned Hamilton structures, the so-called generalized Hamilton spaces~\cite{miron2001geometry}. Given the above progresses, it is nowadays possible to attempt some investigations about the phenomenological and theoretical implications of this framework. Attempts in this direction can for example be found in~\cite{Relancio:2020zok,Relancio:2020rys,Relancio:2020mpa}. 

In this sense, it is quite clear that black holes are particularly interesting objects as their evaporation appears to be a first example of mixing of QFT and GR. However, they are also a source of apparent contradiction between these established theories. One of the most noticeable examples is the so-called information loss problem associated to quantum black hole evaporation~\cite{Hawking:1976ra}. In the GR framework, it seems to imply an non-unitary evolution of quantum states, so violating a basic tenet of QFT. On the other hand, requiring  unitarity seems to imply a breakdown of the equivalence principle, the base of GR, at the event horizon of the black hole (see e.g. the so-called firewall paradigm proposed in~\cite{Almheiri:2012rt}). Therefore, it is natural and pressing to further understand black holes in the above discussed alternative scenarios, where Lorentz invariance can be broken or deformed, to see if these issues survive or are resolved somehow.

In the LIV framework, black holes have been considered in~\cite{Dubovsky:2006vk,Barausse:2011pu,Blas:2011ni,Bhattacharyya:2015gwa}, showing that particles with different energies see different horizons. %Moreover, the fact that generically there is not a common Killing horizon in LIV theories 
This is problematic, gievn that his feature can in principle be used to violate the second law of thermodynamics already at the classical level, as well as via particle dependent Hawking radiation (see e.g.~\cite{Dubovsky:2006vk}), hence pointing towards a basic inconsistency of the LIV scenarios. However, there is evidence that the dynamical realization of such violations of the second law might be practically precluded (see e.g.~\cite{Benkel:2018abt}) or that the UV completion of LIV theories could remedy at this problem by introducing further geometrical structure to the black hole, the so-called Universal Horizon~\cite{Blas:2011ni}. The latter appears to fix a universal temperature and restore black hole thermodynamics~\cite{Herrero-Valea:2020fqa} (see however also~\cite{Michel:2015rsa}). 

In the DSR context, the temperature of black holes has been explored in~\cite{Gim:2014ira,Gim:2015yxa,Mu:2015qna,Kim:2016qtp,Tao:2016baz,Bezerra:2017hrb,Feng:2017iso,Feng:2018jqf,Shahjalal:2018hid}. The main outcome of these investigations is the realization that the the black hole cannot fully evaporate, so ending in a remnant of finite mass. However, in these scenarios the adopted geometrical setting does not include the composition law associated to the deformed kinematics in a clear way, leaving open the question of the compatibility of black hole thermodynamics with momentum dependent geometrical setups fully accounting for the deformed/doubly special relativistic kinematics. 

Fortunately, the aforementioned construction of a cotangent bundle metric can be carried on for a Schwarzschild black hole. In this case, a first remarkable result is that, in spite of the particle-dependent modified dispersion relations, there is still a common horizon for all particles, independently of their energy~\cite{Relancio:2020zok}. However, the usual alternative and equivalent definitions of surface gravity~\cite{Cropp:2013zxi} adopted in GR turn out to not coincide anymore, albeit they can do so in a very particular momentum basis~\cite{Relancio:2021asx}. This opens the question if the laws of black hole mechanics/thermodynamics can be generically extended in this context and in which way.

Following this line of research, in this paper we will hence investigate the first two laws of black hole thermodynamics for a cotangent bundle metric compatible with DSR kinematics. We start by introducing cotangent bundle geometries and recalling some previous results in Sec.~\ref{sec:geometrical_intro}. In Sec.~\ref{sec:zero} and \ref{sec:first} we describe the zero and first laws, respectively, recovering (albeit in a subtle way) the same results of GR. Finally, we discuss the open issue and perspectives in Sec.~\ref{sec:conclusions}. 

\section{The cotangent bundle geometries framework for DSR}
\label{sec:geometrical_intro}

Let us start by laying down the basic building blocks for our analysis. We shall start by introducing the main geometrical ingredients in the cotangent bundle, show their connection to DSR, and finally review our main results from previous papers about their generalization to curved spacetimes.  

\subsection{Cotangent bundle geometries in a nutshell}

In~\cite{miron2001geometry} a line element in the cotangent bundle is defined as  
\begin{equation}
\mathcal{G}\,=\, g_{\mu\nu}(x,k) dx^\mu dx^\nu+g^{\mu\nu}(x,k) \delta k_\mu \delta k_\nu\,,
\label{eq:line_element_ps} 
\end{equation}
where 
\begin{equation}
\delta k_\mu \,=\, d k_\mu - N_{\nu\mu}(x,k)\,dx^\nu\,,
\end{equation}
being  $N_{\nu\mu}(x,k)$ the nonlinear connection coefficients. 

In~\cite{miron2001geometry} it is shown that a horizontal curve is determined by the geodesic motion in spacetime given by
\begin{equation}
\frac{d^2x^\mu}{d\tau^2}+{H^\mu}_{\nu\sigma}(x,k)\frac{dx^\nu}{d\tau}\frac{dx^\sigma}{d\tau}\,=\,0\,,
\label{eq:horizontal_geodesics_curve_definition}
\end{equation} 
and the change of momentum obtained from
\begin{equation}
\frac{\delta k_\lambda}{d \tau}\,=\,\frac{dk_\lambda}{d\tau}-N_{\sigma\lambda} (x,k)\frac{dx^\sigma}{d\tau}\,=\,0\,,
\label{eq:horizontal_momenta}
\end{equation} 
where 
\begin{equation}
{H^\rho}_{\mu\nu}(x,k)\,=\,\frac{1}{2}g^{\rho\sigma}(x,k)\left(\frac{\delta g_{\sigma\nu}(x,k)}{\delta x^\mu} +\frac{\delta g_{\sigma\mu}(x,k)}{\delta  x^\nu} -\frac{\delta g_{\mu\nu}(x,k)}{\delta x^\sigma} \right)\,,
\label{eq:affine_connection_st}
\end{equation} 
is the affine connection of spacetime, and 
\begin{equation}
\frac{\delta}{\delta x^\mu}\, \doteq \,\frac{\partial}{\partial x^\mu}+N_{\nu\mu}(x,k)\frac{ \partial}{\partial k_\nu}\,.
\label{eq:delta_derivative}
\end{equation}
Here, $\tau$ plays the role of the proper time or the affine parametrization depending if one is considering a massive or a massless particle, respectively. 

The choice of the nonlinear connection coefficients $N_{\nu\mu}(x,k)$ is not unique but, as it is shown in~\cite{miron2001geometry}, there is one and only one choice of nonlinear connection coefficients that leads to a space-time affine connection which is metric compatible and torsion free.
When the space-time affine connection is momentum independent, which does not necessarily implies that the momentum metric must also show such behavior,  the coefficients of the nonlinear connection are given by
\begin{equation}
N_{\mu\nu}(x,k)\, = \, k_\rho {\Gamma^\rho}_{\mu\nu}(x)\,,
\label{eq:nonlinear_connection}
\end{equation} 
where $ \Gamma^\rho_{\mu\nu}(x)$ is the GR affine connection. From here it is easy to see that,  when the metric does not depend on the space-time coordinates, these coefficients vanish.

In~\cite{miron2001geometry} the covariant derivatives in space-time were defined as
\begin{equation}
\begin{split}
&T^{\alpha_1 \ldots\alpha_r}_{\beta_1\ldots\beta_s;\mu}(x,k)\,=\,\frac{\delta T^{\alpha_1 \ldots\alpha_r}_{\beta_1\ldots\beta_s}(x,k)}{\delta x^\mu}+T^{\lambda \alpha_2 \ldots\alpha_r}_{\beta_1\ldots\beta_s}(x,k){H^{\alpha_1}}_{\lambda \mu}(x,k)+\cdots+\\
&T^{\alpha_1 \ldots \lambda}_{\beta_1\ldots\beta_s}(x,k){H^{\alpha_r}}_{\lambda \mu}(x,k)-T^{\alpha_1 \ldots \alpha_r}_{\lambda \beta_2\ldots\beta_s}(x,k){H^{\lambda}}_{\beta_1 \mu}(x,k)-\cdots-T^{\alpha_1 \ldots \alpha_r}_{\beta_1\ldots \lambda}(x,k){H^{\lambda}}_{\beta_s \mu}(x,k)\,.
\label{eq:cov_dev_st}
\end{split}
\end{equation} 
Also, it can be shown~\cite{miron2001geometry} that, given a metric, there is always a symmetric non-linear connection leading to metric compatible affine connections in spacetime so that $g_{\mu\nu;\rho}(x,k)=0$.

In order to study the thermodynamics of black holes, we need to know how the Lie derivative is deformed in this context. In~\cite{Barcaroli:2015xda,Relancio:2020zok} the modified Killing equation for a metric in the cotangent bundle was derived
\begin{equation}
\frac{\partial g_ {\mu\nu}(x,k)}{\partial x^\alpha} \chi^\alpha  -\frac{\partial g_ {\mu\nu}(x,k)}{\partial k_\alpha}\frac{\partial \chi^\gamma}{\partial x^\alpha}k_\gamma + g_{\alpha\nu}(x,k)\frac{\partial\chi^\alpha}{\partial x^\mu}+ g_{\alpha\mu}(x,k)\frac{\partial\chi^\alpha}{\partial x^\nu}\,=\,0\,,
\label{eq:killing}
\end{equation}
where $\chi^\alpha=\chi^\alpha(x)$ is momentum independent. In GR, where the metric does not depend on the momentum, the previous condition can be written as 
\begin{equation}
\chi_{\mu\,;\nu}+\chi_{\nu\,;\mu}\,=\,0\,.
\label{eq:killing_GR}
\end{equation}

In~\cite{Relancio:2020zok} it was shown that the Killing equation~\eqref{eq:killing} can be written as (for the particular case in which the affine connection and the nonlinear coefficients are related as in Eq.~\eqref{eq:nonlinear_connection})
\begin{equation}
\chi_{\mu\,;\nu}+\chi_{\nu\,;\mu}-\frac{\partial g_{\mu\nu}}{\partial k_\sigma}\chi^\rho_{;\sigma} k_\rho\,=\,0\,.
\label{eq:killing_DGR}
\end{equation}

\subsection{Deformed relativistic kinematics in curves spacetimes}
We summarize here our previous results about supplementing a deformed relativistic kinematics within a curved spacetime. 

The deformed kinematics of DSR are usually obtained from Hopf algebras~\cite{Majid:1995qg}, being the $\kappa$-Poincaré kinematics~\cite{Majid1994}  the most studied example of this kind of construction. This kinematics has been understood from a geometrical point of view in~\cite{Carmona:2019fwf}. Given a de Sitter momentum metric $ \tilde{g}$, translations can be used to define the associative deformed composition law, the Lorentz isometries lead to the Lorentz transformations, and the (squared of the) distance in momentum space is identified with the deformed Casimir.   

In~\cite{Relancio:2020zok,Pfeifer:2021tas}, we extended~\cite{Carmona:2019fwf} in order to consider in the same framework a deformed kinematics and a curved spacetime. For that aim, it is mandatory to consider the cotangent bundle geometry above discussed. The metric tensor $g_{\mu\nu}(x,k)$ in the cotangent bundle depending on space-time coordinates was constructed via the space-time tetrad and the metric in momentum space, $\tilde{g}$, so that
\begin{equation}
g_{\mu\nu}(x,k)\,=\,e^\alpha_\mu(x) \tilde{g}_{\alpha\beta}(\bar{k})e^\beta_\nu(x)\,,
\label{eq:definition_metric_cotangent}
\end{equation}
where $\bar{k}_\alpha=\bar{e}^\nu_\alpha (x) k_\nu$, and $\bar{e}$ denotes the inverse of the space-time tetrad. 

In~\cite{Relancio:2020rys} it was also proved that this Hamiltonian can be identified with the square of the minimal geometric distance of a momentum $k$ from the origin of momentum space, relating  the Casimir and the metric in the following way~\cite{Relancio:2020zok}
\begin{equation}
\mathcal{C}(x,k)\,=\,\frac{1}{4} \frac{\partial \mathcal{C}(x,k)}{\partial k_\mu} g_{\mu\nu} (x,k) \frac{\partial \mathcal{C}(x,k)}{\partial k_\nu}\,.
\label{eq:casimir_metric}
\end{equation} 
It is also possible to choose any function of such Casimir as a new Casimir, which tantamount to make a redefinition of the mass~\cite{Carmona:2019fwf}.

A very important relation satisfied by the Casimir  is that its delta derivative~\eqref{eq:delta_derivative} is zero, i.e.
\begin{equation}
\frac{\delta \mathcal{C}(x,k)}{\delta x^\mu}\,=\,0\,.
\label{eq:casimir_delta}
\end{equation} 
This is a necessary condition derived from the fact the Hamilton equations of motions are horizontal curves~\cite{Relancio:2020rys}. 

In~\cite{Pfeifer:2021tas} we showed that the most general form of the metric, in which the construction of a deformed kinematics in a curved space-time background is allowed, is a momentum basis whose Lorentz isometries are linear transformations in momenta, i.e., a metric of the form
\begin{equation}
\tilde{g}_{\mu\nu}(k)\,=\,\eta_{\mu \nu} f_1 (k^2)+\frac{k_\mu k_\nu}{\Lambda^2} f_2(k^2)\,,
\label{eq:lorentz_metric}
\end{equation}
where $\Lambda$ is the high-energy scale parametrizing the momentum deformation of the metric and kinematics,  $\eta_{\mu\nu}$ is the Minkowski metric, and $k^2=k_\mu \eta^{\mu\nu}k_\nu$ . It is also important to note that, in order to have a smooth limit to SR, 
\begin{equation}
f_1(0)\,=\,1 \,,\qquad f_2(0)\,=\,\text{const}\,.
\label{eq:functions}
\end{equation}

From Eq.~\eqref{eq:definition_metric_cotangent} one obtains the following metric in the cotangent bundle when a curvature in spacetime is present
\begin{equation}
g_{\alpha\beta}(x,k)\,=\,a_{\alpha\beta}(x)f_1 (\bar{k}^2)+\frac{k_\alpha k_\beta}{\Lambda^2} f_2(\bar{k}^2)\,,
\label{eq:lorentz_metric_curved}
\end{equation}
where $a_{\mu \nu}(x) = e^\alpha_\mu(x) \eta_{\alpha\beta} e^\beta_\nu(x)$ is the GR metric, and $\bar k^2=\bar k_\mu \eta^{\mu\nu} \bar k_\nu=k_\mu a^{\mu\nu}(x)  k_\nu$. Therefore, one can use the definition of the space-time affine connection~\eqref{eq:affine_connection_st} to show that this connection is momentum independent, being then $ H^\rho_{\mu\nu}(x,k)=\Gamma^\rho_{\mu\nu}(x)$, the same one of GR~\cite{Pfeifer:2021tas}.

Since we want the momentum metric~\eqref{eq:lorentz_metric} to be a de Sitter space (allowing us to define a deformed relativistic kinematics), a relationship between the functions $f_1$ and $f_2$ must hold. In particular, in~\cite{Relancio:2020rys} it was shown that, in order to be conserved the Einstein's tensor, as defined in~\cite{miron2001geometry} (which is the same expression of GR), the metric~\eqref{eq:lorentz_metric_curved} must be conformally flat. Then, taking $f_2=0$ and imposing a momentum de Sitter space, one obtains from Eq.~\eqref{eq:lorentz_metric_curved} 
\begin{equation}
g_{\mu\nu}(x,k)\,=\,a_{\mu \nu}(x)\left(1-\frac{\bar{k}^2}{4\Lambda^2}\right)^2\,.
\label{eq:conformal_metric}
\end{equation}

\subsection{Black holes in the cotangent bundle}
\label{subsec:BH}

Following the above definitions, it is easy to see that the deformed Kerr metric in the Boyer-Lindquist coordinates reads
 \begin{equation}
 \begin{split}
ds^2\,=\,& f_1(\bar k^2)\left(-\left(1-\frac{2 M  r}{\rho^2}\right) d v^2+2 d v\,d r-2 a \sin^2  \theta \, d r\, d\psi -\frac{4 M a  r \sin^2  \theta }{\rho^2} d v \, d \psi \right.\\
&\left. +\frac{\Sigma }{\rho^2} \sin^2  \theta\,  d\psi^2 +\rho^2 d \theta^2\right) +\frac{f_2(\bar k^2)}{\Lambda^2}\left(k_v dv +k_r dr+k_\theta d\theta+k_\psi d\psi \right)^2 \,,
 \end{split}
\label{eq:line_kerr}
\end{equation} 
where  
\begin{equation}
\Sigma\,=\, \left( r^2+a^2\right)^2-a^2 \Delta  \sin^2  \theta\,,\qquad \Delta\,=\,  r^2-2M  r-a^2\,,\qquad \rho^2\,=\,  r^2+a^2  \cos^2  \theta\,.
\end{equation} 

In~\cite{Relancio:2021asx} it was found that only for the basis leading to Eq.~\eqref{eq:conformal_metric} the diverse possible notions of surface gravity coincide. While this is always true in GR when there is a Killing horizon~\cite{Cropp:2013zxi}, this is not longer true in these generic cotangent bundle manifolds. However, for the basis of Eq.~\eqref{eq:conformal_metric}, in contrast with any other momentum coordinates, the Killing equation can be written as in GR, i.e., Eq.~\eqref{eq:killing_GR}, allowing for the usual derivation of surface gravity definition to go trough.

In spite of this, we shall try to keep our analysis as general as possible, and hence, in what follows, we will use the general form Eq.~\eqref{eq:lorentz_metric_curved}, with generic functions $f_1$ and $f_2$.

\section{Zero law}
\label{sec:zero}
As we saw in~\cite{Relancio:2021asx}, the surface gravity obtained from the inaffinity of null geodesics is the same of the one of GR for a generic metric of the form of~\eqref{eq:lorentz_metric_curved}, i.e., independently of the choice of $f_1$ and $f_2$. Here we show that, for any choice of these functions, the surface gravity is a constant, as in GR, when considering massless particles.  For that aim, we start by showing that the Killing energy is still a conserved quantity. From the aforementioned definition of the surface gravity we are able to prove that the 
 ``generator'' surface gravity~\cite{Cropp:2013zxi}, another common (and useful) notion, is the same one of GR, and hence coincides with $\kappa_\text{inaffinity}$, independently again on the functions $f_1$ and $f_2$. Given that, it is easy to prove that the zero law is still valid in this scheme. 

\subsection{Conserved Killing energy}

We start by considering the derivative of the Killing energy with respect the affine parameter $\tau$
 \begin{equation}
 \frac{d (u^\alpha \chi_\alpha)}{d\tau}\,=\,  \left(u^\alpha \chi_\alpha\right)_{;\mu} u^\mu\,,
\label{eq:killing_energy}
\end{equation}
where $u^\alpha$ is the velocity vector. As showed in~\cite{Relancio:2020rys}, the derivative with respect to $\tau$ can be expressed also in this cotangent bundle framework as a covariant derivative contracted with the velocity. For our case study we will consider only massless particles, so $u^2=0$. This also means that, due to the linear Lorentz covariant form of the metric~\eqref{eq:lorentz_metric_curved}, the same relation between momentum and velocity of GR holds in this scheme ($k_\mu=a_{\mu\nu}u^\nu$), and therefore
 \begin{equation}
k_\mu g^{\mu\nu} k_\nu\,=\,k_\mu a^{\mu\nu} k_\nu\,=\,k_\mu u^\mu\,=\,0\,,
\label{eq:v_k_null}
\end{equation}
since $\bar k^2 =0$.

We can now expand Eq.~\eqref{eq:killing_energy} as
\begin{equation}
u^\alpha_{;\mu} u^\mu \chi_\alpha+u^\alpha u^\mu \chi_{\alpha;\mu}\,=\, \frac{1}{2}\frac{\partial g_{\mu\alpha}}{\partial k_\sigma}\chi^\rho_{;\sigma} k_\rho  u^\alpha u^\mu \,=\, - g_{\mu\alpha}\frac{\partial  u^\alpha}{\partial k_\sigma}  u^\mu  {\chi^\rho}_{;\sigma} k_\rho\,=\,0\,,
\label{eq:killing_energy2}
\end{equation}
where in the second step we have used the geodesic equation and the Killing equation~\eqref{eq:killing_DGR}, in the third one the fact that $u^2=0$, and in the last one the fact that the four-velocity for massless particles does not depend on $k$, since  they probe a momentum independent geometry as it can be seen by contracting~\eqref{eq:lorentz_metric_curved} twice with the momentum $k^\mu$. With this we show that, for massless particles, there is a conserved quantity along a geodesic, which is the usual Killing energy, and that the following relation holds
 \begin{equation}
 u^\alpha u^\mu \chi_{\alpha;\mu}\,=\,0\,,
\label{eq:killing_vector_v}
\end{equation}
which will be useful in the following. 

\subsection{Generator surface gravity}
It was shown in~\cite{Relancio:2021asx} that the surface gravity, defined as the inaffinity of the generators of the horizon, is the same as in GR for a black hole metric of the from~\eqref{eq:lorentz_metric_curved}, i.e.
 \begin{equation}
 \chi^\nu\,  {\chi^\mu}_{;\nu} \,=\,\kappa_{\text{inaffinity}} \chi^\mu\,.
\label{eq:inaffinity}
\end{equation}
This is due to the fact that, if the  Killing vector is a constant in the GR metric, it will be still constant also in our framework. This can be understood just by looking at Eq.~\eqref{eq:killing}. Therefore, the surface gravity defined in Eq.~\eqref{eq:inaffinity} will be the same one of GR, since the affine connection is also momentum independent.

We now follow the procedure given in~\cite{Wald:1984rg}, using the Frobenius' theorem~\cite{Poisson:2009pwt}, which states that if a congruence of geodesics has a vanishing rotation tensor, the Killing vector field $\chi_\alpha$ satisfies
 \begin{equation}
 \chi_{[\alpha;\beta} \chi_{\gamma]}\,=\,0\,,
\label{eq:frobenius}
\end{equation}
where the brackets denote indexes antisymmetrization.  Using the Killing equation~\eqref{eq:killing_DGR} we can write the previous expression as
 \begin{equation}
 2 \left( \chi_{\alpha;\beta}\chi_\gamma +\chi_{\gamma;\alpha}\chi_\beta +\chi_{\beta;\gamma}\chi_\alpha \right)\,=\, {\chi^\epsilon}_{;\delta}k_\epsilon\left( \frac{\partial g_{\alpha\beta}}{\partial k_\delta} \chi_\gamma+ \frac{\partial g_{\alpha\gamma}}{\partial k_\delta}  \chi_\beta+ \frac{\partial g_{\beta\gamma}}{\partial k_\delta} \chi_\alpha\right)\,.
\label{eq:frobenius2}
\end{equation}
Multiplying the previous equation by $\chi^{\alpha;\beta}$ we find
 \begin{equation}
 2 \left( \chi_{\alpha;\beta}\chi^{\alpha;\beta}\chi_\gamma+2\kappa^2\chi_\gamma \right)\,=\, {\chi^\epsilon}_{;\delta}k_\epsilon\left( \frac{\partial g_{\alpha\beta}}{\partial k_\delta} \chi_\gamma \chi^{\alpha;\beta}+ 2\kappa \frac{\partial g_{\alpha\gamma}}{\partial k_\delta}  \chi^\alpha  \right)\,,
\label{eq:frobenius3}
\end{equation}
where we have used again~\eqref{eq:killing_DGR}. In order to find a simple relation between the surface gravity and the Killing vector, we can see the explicit form of the derivative of the metric~\eqref{eq:lorentz_metric_curved} with respect the momentum 
 \begin{equation}
\frac{\partial g_{\alpha\beta}}{\partial k_\delta}\,=\, 2 \frac{k_\rho a^{\rho\delta}}{\Lambda^2} \left(a_{\alpha\beta}\frac{\partial f_1}{\partial \bar k^2}+\frac{k_\alpha k_\beta}{\Lambda^2}\frac{\partial f_2}{\partial \bar k^2}\right)+\frac{f_2}{\Lambda^2}\left(k_\alpha \delta^\delta_\beta+k_\beta \delta^\delta_\alpha\right)\,.
\label{eq:dg_dk}
\end{equation}
Using Eq.~\eqref{eq:killing_vector_v} we can see that, when substituting the previous equation into Eq.~\eqref{eq:frobenius3}, the first term of Eq.~\eqref{eq:dg_dk} vanishes. Therefore, we focus on the terms of the right hand side of Eq.~\eqref{eq:frobenius3} when substituting the second term of~\eqref{eq:dg_dk}. Indeed we find
 \begin{equation}
\begin{split}
{\chi^\epsilon}_{;\delta} k_\epsilon &\left( \frac{\partial g_{\alpha\beta}}{\partial k_\delta} \chi_\gamma \chi^{\alpha;\beta}+ 2\kappa \frac{\partial g_{\alpha\gamma}}{\partial k_\delta}  \chi^\alpha  \right)\,=\,k_\epsilon \frac{f_2}{\Lambda^2}\left(k_\alpha \chi_\gamma {\chi^\epsilon}_{;\delta}  \left( \chi^{\alpha;\delta}+ \chi^{\delta;\alpha}\right)\right. \\ &\left.+2 \kappa \left(k_\gamma  \chi^\delta {\chi^\epsilon}_{;\delta}  +k_\alpha \chi^\alpha {\chi^\epsilon}_{;\gamma}  \right)\right)
\,=\,k_\epsilon \frac{f_2}{\Lambda^2}\left(k_\alpha \chi_\gamma {\chi^\epsilon}_{;\delta}  \frac{\partial g^{\alpha \delta}}{k_\rho} {\chi^{\sigma}}_{;\rho}k_\sigma\right.\\
&\left.+2 \kappa \left(k_\gamma \kappa \chi^\epsilon +k_\alpha \chi^\alpha {\chi^\epsilon}_{;\gamma}  \right)\right)\,=\, 2\kappa\frac{f_2}{\Lambda^2}    k_\epsilon\left(k_\gamma \kappa \chi^\epsilon +k_\alpha \chi^\alpha {\chi^\epsilon}_{;\gamma}  \right) \,,
\label{eq:frobenius4}
\end{split}
\end{equation}
where we have used  Eqs.~\eqref{eq:killing_DGR} and \eqref{eq:inaffinity}, and in the last one Eq.~\eqref{eq:killing_vector_v}.
Hence, we can now write Eq.~\eqref{eq:frobenius3} as 
 \begin{equation}
 2 \left( \chi_{\alpha;\beta}\chi^{\alpha;\beta}\chi_\gamma+2\kappa^2\chi_\gamma \right)\,=\, 2\kappa\frac{f_2}{\Lambda^2}    k_\epsilon\left(k_\gamma \kappa \chi^\epsilon +k_\alpha \chi^\alpha {\chi^\epsilon}_{;\gamma}   \right)\,.
\label{eq:frobenius5}
\end{equation}
When multiplying the previous equation by $k^\gamma$ we find that the right hand side vanishes by virtue of Eqs.~\eqref{eq:v_k_null} and \eqref{eq:killing_vector_v}, so we recover
 \begin{equation}
 \kappa^2 \,=\,- \frac{1}{2}  \chi_{\alpha;\beta}\chi^{\alpha;\beta} \,.
\label{eq:frobenius_gr}
\end{equation}
In conclusion, we have proved that for a generic metric of the form~\eqref{eq:lorentz_metric_curved}, the surface gravity defined from the horizon's null generators, can be written as in GR. 

Actually, we can also show that this is the very same expression of GR. We start by writing the inverse of the metric~\eqref{eq:lorentz_metric_curved}
\begin{equation}
g^{\alpha\beta}(x,k)\,=\,a^{\alpha\beta}(x)\frac{1}{f_1 (\bar{k}^2)}-a^{\delta\alpha}(x) a^{\gamma\beta}(x) k_\delta k_\gamma\frac{ f_2(\bar{k}^2)}{ f_1(\bar{k}^2)\left( f_1(\bar{k}^2) \Lambda^2 + f_2(\bar{k}^2)\bar{k}^2\right)}\,.
\label{eq:lorentz_metric_curved_inv}
\end{equation}
Now we can use the metric~\eqref{eq:lorentz_metric_curved} and its inverse~\eqref{eq:lorentz_metric_curved_inv} to rewrite the right hand side of Eq.~\eqref{eq:frobenius_gr} as
 \begin{equation}
 \begin{split}
 {\chi^\alpha}_{;\beta}{\chi^\gamma}_{;\delta}g_{\alpha\gamma}&g^{\beta\delta}\,=\,  {\chi^\alpha}_{;\beta}{\chi^\gamma}_{;\delta}\left(a_{\alpha\gamma}a^{\delta\beta}+\frac{f_2}{f_1\Lambda^2}k_\alpha k_\gamma a^{\beta\delta}\right.\\
 &\left.-a_{\alpha\gamma} a^{\delta\sigma}  a^{\rho\beta} k_\sigma k_\rho\frac{ f_2 }{ f_1  \Lambda^2 + f_2 \bar{k}^2}- a^{\delta\sigma}  a^{\rho\beta} k_\sigma k_\rho k_\alpha k_\gamma \frac{ f_2^2 }{f_1\left( f_1  \Lambda^2 + f_2 \bar{k}^2\right)}\right) \,.
\label{eq:frobenius_gr2}
 \end{split}
\end{equation}
The last term vanishes by virtue of Eq.~\eqref{eq:killing_vector_v}. Moreover, when considering massless particles, $\bar k^2=0$, so the coefficients of the second and third terms are the same. By taking into account Eq.~\eqref{eq:killing_GR}, which is valid for the GR metric $a_{\mu \nu}$, one can finally obtain
 \begin{equation}
 \kappa^2 \,=\,- \frac{1}{2}   {\chi^\alpha}_{;\beta}{\chi^\gamma}_{;\delta}a_{\alpha\gamma}a^{\beta\delta} \,,
\label{eq:frobenius_gr3}
\end{equation}
which is exactly the same momentum independent GR expression, which we already know to be equivalent to $\kappa_\text{inaffinity}$.

\subsection{Uniformity of the surface gravity on the horizon}
Strong of the above results, it is now trivial to prove that the surface gravity is uniform on the event horizon even in our cotangent bundle setting. Since we have seen that the surface gravity defined from null generators~\eqref{eq:frobenius_gr3} is momentum independent, we can follow the usual GR procedure. As showed in~\cite{Wald:1984rg}, from Eq.~\eqref{eq:killing_GR} one can obtain 
 \begin{equation}
{\chi^\alpha}_{;\mu;\nu} \,=\,{R^\alpha}_{\mu\nu\sigma}(x)\chi^\sigma \,,
\label{eq:2cov_dev}
\end{equation}
where ${R^\sigma}_{\alpha\mu\nu}(x)$ is the Riemann's tensor~\footnote{Note that, while in a generic cotangent bundle metric the Riemann tensor, defined from the commutator of two covariant derivatives, depends also on the momentum, this is not the case for our particular metric~\eqref{eq:lorentz_metric_curved}, since the affine connection is the same one of GR. Moreover, since the Killing vector is momentum independent, there are not extra (momentum dependent) terms that should be taken into account in~\eqref{eq:2cov_dev}~\cite{miron2001geometry}}. Using this expression, we can see from Eq.~\eqref{eq:frobenius_gr3}
 \begin{equation}
 \chi^{\sigma}\kappa^2_{;\sigma} \,=\, 2 \kappa \chi^{\sigma}\kappa_{;\sigma} \,=\,-  {\chi^\alpha}_{;\beta}a_{\alpha\gamma}a^{\beta\delta} {R^\gamma}_{\delta\rho\sigma}\chi^\rho \chi^{\sigma}\,=\,0\,,
\label{eq:kappa_der1}
\end{equation}
due to the symmetry of the Riemann's tensor. Similarly one can also prove that along a transverse direction the surface gravity is a conserved quantity, as in~\cite{Wald:1984rg}.

Therefore, in this section we have proved that the zero law of black hole thermodynamics is also valid in this scheme, independently of the choice of the function $f_1$ and $f_2$ defining the momentum metric.   

\section{  Smarr's formula and first law}
\label{sec:first}
In this section we discuss the Smarr's formula and the first law of black hole thermodynamics, which can be derived from the former. In particular we show that, by taking into account that the physical distance measured by the particle {probing the spacetime} is momentum dependent, one finds no modification with respect to the GR scenario. 

\subsection{Smarr's formula}
We start by regarding the space-time line element from Eq.~\eqref{eq:line_element_ps}, which is
  \begin{equation}
ds^2\,=\, dx^\mu g_{\mu\nu}(x,k)dx^\nu\,.
\label{eq:line_element_st}
\end{equation} 
In order to take into account the momentum dependent effects when regarding this line element, we must substitute momenta as a function of the space-time coordinates~\footnote{Through a space-time geodesic (horizontal curve) momenta satisfy Eq.~\eqref{eq:horizontal_momenta}, being then a function of the velocity and the space-time coordinates. In particular, there is a very simple way to find this dependence: since the velocity can be obtained as a function of the coordinates by solving the geodesic equation~\eqref{eq:horizontal_geodesics_curve_definition} (remember that in our framework the space-time affine connection is momentum independent), momenta can be written as the same function times the GR metric, due to the fact that in this scenario, $k_\mu =a_{\mu \nu}(x)v^\nu$ for massless particles.}. Therefore, the line element becomes 
  \begin{equation}
ds^2\,=\, dx^\mu g_{\mu\nu}(x,k(x))dx^\nu\,.
\label{eq:line_element_st2}
\end{equation} 
Since we are interested in this section in computing the area as an integral, using the space-time line element, we can define new  effective space-time coordinates, $\tilde x^\mu$~\cite{Relancio:2020mpa}, in such a way that the previous line element can be written as 
  \begin{equation}
ds^2\,=\, d\tilde{x}^\mu a_{\mu\nu}(\tilde{x})d\tilde{x}^\nu\,,
\label{eq:line_element_st_t}
\end{equation} 
which is the same line element of GR but in these new tilde coordinates. In this way we are assuming that the physical distance traveled by a particle, through a momentum dependent geometry, is also momentum dependent. 

Note that this change of variables cannot be made in general since the metric is not invariant under change of momentum coordinates~\cite{Relancio:2020rys}. However, since we are only interested in the measure on spacetime, we can freely use this definition. In particular, if the momentum metric can be written as a function of a tetrad,
\begin{equation}
    \tilde g_{\mu\nu}(\bar k)\,=\,\varphi^\alpha_\mu(\bar k) \eta_{\alpha\beta} \varphi^\beta_\nu(\bar k)\,, 
\end{equation}
the new tilde coordinates can be expressed through the following relationship
\begin{equation}
    d  \tilde x^\mu e^\alpha_\mu (\tilde x)\,=\,   d x^\mu e^\beta_\mu (x)\varphi^\alpha_\beta(\bar k)\,. 
\end{equation}

This allows us to rederive the Smarr's formula in a simple and direct way. We start by considering the area of the horizon of a black hole.  We use the following coordinates of the Kerr's black hole~\cite{Poisson:2009pwt}
 \begin{equation}
 \begin{split}
ds^2\,=\,& -\left(1-\frac{2 M \tilde r}{\tilde\rho^2}\right) d\tilde v^2+2 d\tilde v\,d\tilde r-2 a \sin^2  \tilde\theta \, d\tilde r\, d\tilde\psi -\frac{4 M a \tilde r \sin^2  \tilde\theta }{\tilde\rho^2} d\tilde v \, d\tilde \psi \\
&+\frac{\tilde\Sigma }{\tilde\rho^2} \sin^2 \tilde \theta\,  d\psi^2 +\rho^2 d\tilde \theta^2\,,
\label{eq:line_kerr_tilde}
 \end{split}
\end{equation} 
where,
 \begin{equation}
\tilde\Sigma\,=\, \left(\tilde r^2+a^2\right)^2-a^2 \tilde\Delta  \sin^2 \tilde \theta\,,\qquad \tilde\Delta\,=\, \tilde r^2-2M \tilde r-a^2\,,\qquad \tilde\rho^2\,=\, \tilde r^2+a^2  \cos^2  \tilde\theta\,.
\end{equation} 
The area is defined as 
 \begin{equation}
A\,=\, \oint_\mathcal{H} \sqrt{\tilde\sigma} d^2\tilde\theta \,,
\label{eq:area}
\end{equation} 
 where $\mathcal{H}$ is a two-dimensional cross section of the event horizon, for $\tilde v$ constant, $\tilde r=\tilde r_+=r_+=M+\sqrt{M^2-a^2}$, $0\leq\tilde \theta\leq \pi$, and  $0\leq\tilde\psi<2\pi$. The equality $\tilde r_+= r_+$ holds because the event horizon is the same for all massless particles, independently of their energy~\cite{Relancio:2020zok}. This can be easily understood from the fact that the geodesics of massless particles are undeformed for the metric~\eqref{eq:lorentz_metric_curved}, as explained previously. 

Using~\eqref{eq:line_kerr_tilde} one  finds
 \begin{equation}
\sqrt{\tilde\sigma}\,=\, \sqrt{  \tilde\Sigma \sin^2 \tilde \theta } \,,
\end{equation} 
so one obtains from Eq.~\eqref{eq:area} 
 \begin{equation}
A\,=\, 4 \pi\left(r_+^2+a^2\right)\,.
\label{eq:area2}
\end{equation} 
In order to derive the Smarr's formula we start by multiplying the area by the surface gravity $\kappa=\sqrt{M^2-a^2}/\left(r_+^2+a^2\right)$~\footnote{Remember that in the previous section we have shown that the same GR surface gravity is also obtained in this cotangent bundle framework from the inaffinity and generator definition.}, and dividing it by $4\pi$, obtaining
 \begin{equation}
\frac{\kappa A}{4 \pi}\,=\, \left(r_+^2+a^2\right)\frac{\sqrt{M^2-a^2}}{r_+^2+a^2}\,=\, \sqrt{M^2-a^2}\,.
\label{eq:smarr1}
\end{equation} 
Now, we can multiply  the angular velocity of the black hole, which is  $\Omega_H=a/(r_+^2+a^2)$, by  $J=M a$, the angular momentum of the black hole, finding the following expression 
 \begin{equation}
2 \Omega J\,=\, \frac{2 a^2 M}{r_+^2+a^2}\,.
\label{eq:smarr2}
\end{equation} 
Adding Eqs.~\eqref{eq:smarr1} and~\eqref{eq:smarr2}, and using that the explicit value of the event horizon radius, $r_+=M+\sqrt{M^2-a^2}$, one can finally obtain the well-known result
 \begin{equation}
M\,=\, 2 \Omega_H J+\frac{\kappa A}{4 \pi}\,,
\label{eq:smarr}
\end{equation} 
which is the same GR expression of the Smarr's formula.

\subsection{First law}
From Eq.~\eqref{eq:smarr} one is able to derive the first law of black hole thermodynamics~\cite{Poisson:2009pwt}. Let us consider a black hole with mass $M$ and angular momentum $J$ that is perturbed, so these parameters become $M+\delta M$ and $J+\delta J$.  We can see that both variations are related by deriving Eq.~\eqref{eq:area2}
 \begin{equation}
\frac{\delta A}{8 \pi}\,=\,  r_+ \delta r_+ + a \delta a\,.
\label{eq:variation_area}
\end{equation} 
By taking into account the relationship between $r_+$ and the mass and spin of the black hole we find 
 \begin{equation}
\left(r_+-M\right) \delta r_+\,=\,  r_+  \delta M- a \delta a\,,
\label{eq:variation_area2}
\end{equation} 
and from here it is easy to verify that
 \begin{equation}
\frac{\kappa}{8 \pi} \delta A \,=\, \delta M- \Omega_H \delta J\,,
\label{eq:variation_area3}
\end{equation} 
holds, exactly as in GR.

\section{Conclusion}
\label{sec:conclusions}
In this paper we have addressed, for the first time, the first two black hole laws of thermodynamics in a cotangent bundle geometry associated to the deformed kinematics of doubly special relativity. We have found that these laws are unmodified with respect to the general relativistic setting for the most general metric in the cotangent bundle compatible with a consistent lift of the deformed symmetries on curved spacetimes, Eq.~\eqref{eq:lorentz_metric_curved}.

The zeroth law implies the conservation of the surface gravity, being it a constant for an stationary black hole. In a previous work~\cite{Relancio:2021asx} we have seen that not every notion of surface gravity is equivalent in this scheme, being all of them equal for a particular choice of the momentum metric Eq.~\eqref{eq:conformal_metric}. Combining the results of that paper with the ones presented here, we see that both notions related to inaffinity of geodesics, peeling of null geodesics, and generator of null horizon, coincide for any metric of the form Eq.~\eqref{eq:lorentz_metric_curved}, being them the same one of general relativity. These definitions seems to be more fundamental in our framework as they are valid and equivalent even for the most general metric~\eqref{eq:lorentz_metric_curved}, while the other definitions only agree for the metric displayed in~\eqref{eq:conformal_metric}.

Regarding the first law, we have seen that, taking into account that the physical distance measured by particles is momentum dependent, it leads also to the same result of general relativity. This implies that the same relationship between mass, angular momentum, and area of a black hole, is independent on the particle probing spacetime. 

The extension of the remaining two laws of black hole mechanics is far from trivial given that the second law would require a more detailed understanding of the field equations, while the third one would require an in dept analysis of trapped surfaces in our framework. Let us also stress that the recovery of these two laws of black hole mechanics in their usual GR form strongly suggests that the surface gravity will determine the quantum black hole temperature in our framework, as in the usual case. However, a confirmation of this conjecture can only be achieved by a suitable extension of the standard derivation of Hawking radiation based on QFT in curved spacetime to the full cotangent bundle geometry. 
We hope to be able to address these open issues in future works.

\section*{Acknowledgments}

JJR acknowledges support from the INFN Iniziativa Specifica GeoSymQFT. SL acknowledge funding from the Italian Ministry of Education and Scientific Research (MIUR) under the grant PRIN MIUR 2017-MB8AEZ. We thank Christian Pfeifer for useful comments.  The authors would also like to thank support from the COST Action CA18108.

\end{document}